%% file: paper.tex
\documentclass[runningheads]{llncs}
\usepackage{graphicx}
\usepackage{amsmath}
\usepackage{bm}
\DeclareMathOperator*{\argmin}{arg\,min}
\usepackage{float}
\usepackage{amsfonts}
\usepackage{booktabs}
\usepackage{siunitx}
\usepackage{hyperref}
\hypersetup{
  colorlinks   = true, 
  urlcolor     = blue, 
  linkcolor    = blue, 
  citecolor   = blue 
}
\usepackage{multirow}
\usepackage{tabularx}
\usepackage{subfigure}

\usepackage{algorithm2e}
\RestyleAlgo{ruled}
\SetAlFnt{\tiny}
\SetAlCapFnt{\scriptsize}
\SetAlCapNameFnt{\scriptsize}

\begin{document}
\title{Deep Learning-based Anonymization of Chest Radiographs: A Utility-preserving Measure for Patient Privacy}
\titlerunning{Deep Learning-based Anonymization of Chest Radiographs}
%
\author{Kai Packhäuser \and Sebastian Gündel \and Florian Thamm \and Felix Denzinger \and Andreas Maier}
%
\authorrunning{K. Packhäuser et al.}
%
\institute{Friedrich-Alexander-Universität Erlangen-Nürnberg, Erlangen, Germany \\
\email{kai.packhaeuser@fau.de}}
\maketitle              
\begin{abstract}
Robust and reliable anonymization of chest radiographs constitutes an essential step before publishing large datasets of such for research purposes. The conventional anonymization process is carried out by obscuring personal information in the images with black boxes and removing or replacing meta-information. However, such simple measures retain biometric information in the chest radiographs, allowing patients to be re-identified by a linkage attack. Therefore, there is an urgent need to obfuscate the biometric information appearing in the images. We propose the first deep learning-based approach (PriCheXy-Net) to targetedly anonymize chest radiographs while maintaining data utility for diagnostic and machine learning purposes. Our model architecture is a composition of three independent neural networks that, when collectively used, allow for learning a deformation field that is able to impede patient re-identification. Quantitative results on the ChestX-ray14 dataset show a reduction of patient re-identification from~81.8\,\% to~57.7\,\%~(AUC) after re-training with little impact on the abnormality classification performance. This indicates the ability to preserve underlying abnormality patterns while increasing patient privacy. Lastly, we compare our proposed anonymization approach with two other obfuscation-based methods (Privacy-Net, DP-Pix) and demonstrate the superiority of our method towards resolving the privacy-utility trade-off for chest radiographs.

\end{abstract}

\section{Introduction}
Deep learning~(DL)~\cite{lecun2015deep} has positively contributed to the development of diagnostic algorithms for the detection and classification of lung diseases in recent years~\cite{guendel2018learning,maier2019gentle}. This progress can largely be attributed to the availability of public chest X-ray datasets~\cite{bustos2020padchest,cohen2020covid,irvin2019chexpert,johnson2019mimic,wang2017chestx}. However, as chest radiographs inherently contain biometric information (similar to a fingerprint), the public release of such data bears the risk of automated patient re-identification by DL-based linkage attacks~\cite{packhauser2022deep}. This would allow patient-related information, e.\,g., age, gender, or disease findings to be revealed. Therefore, there is an urgent need for stronger privacy mechanisms for chest X-ray data to alleviate the risk of linkage attacks.

Perturbation-based anonymization approaches~\cite{kaissis2020secure}~--~such as differential privacy~(DP)~\cite{dwork2008differential,dwork2006calibrating}~-- have become the gold standard to obfuscate biometric identifiers from sensitive data. Such approaches are based on the postulate that the global statistical distribution of a dataset is retained while personal information is reduced. This can be realized by applying randomized modifications, i.\,e. noise, to either the inputs~\cite{bu2020deep,dong2019gaussian,dwork2006calibrating}, the computational results, or to algorithms~\cite{abadi2016deep,ziller2021medical}. Although originally proposed for statistical data, DP has been extended to image data with the differentially private pixelization method~(DP-Pix)~\cite{fan2018image,fan2019differential}. This method involves pixelizing an image by averaging the pixel values of each $b\times b$ grid cell, followed by adding Laplace noise with 0 mean and $\frac{255m}{b^2\epsilon}$ scale. Parameter~$\epsilon$ is used to determine the privacy budget (smaller values indicate greater privacy), while the $m$-neighborhood represents a sensitivity factor. However, one major drawback of perturbation-based anonymization is the potential degradation of image quality and an associated reduction in data utility.

In recent years, DL has emerged as a prominent tool for anonymizing medical images. In this context, synthetic image generation with privacy guarantees is currently actively explored, aimed at creating fully anonymous medical image datasets~\cite{packhauser2022generation,pinaya2022brain}. Furthermore, adversarial approaches -- such as Privacy-Net~\cite{kim2021privacy} -- have been proposed, which focus on concealing biometric information while maintaining data utility. Privacy-Net is composed of a U-Net encoder that predicts an anonymized image, an identity discriminator, and a task-specific segmentation network. Through the dual process of deceiving the discriminator and optimizing the segmentation network, the encoder acquires the ability to obfuscate patient-specific patterns, while preserving those necessary for the downstream task. However, while originally designed for utility preservation on MRI segmentation tasks, the direct applicability and transferability of Privacy-Net to other image modalities and downstream tasks (e.\,g. chest X-ray classification) is limited. As we will experimentally demonstrate in our study, more sophisticated constraints are required for the utility-preserving anonymization of chest radiographs in order to successfully maintain fine-grained abnormality details that are crucial for reliable classification tasks.

In this work, we aim to resolve the privacy-utility trade-off by proposing the~-- to the best of our knowledge -- first adversarial image anonymization approach for chest radiography data. Our proposed model architecture~(PriCheXy-Net) is a composition of three independent neural networks that collectively allow for the learning of targeted image deformations to deceive a well-trained patient verification model. We apply our method to the publicly available ChestX-ray14 dataset~\cite{wang2017chestx} and evaluate the impact of different deformation degrees on anonymization capability and utility preservation. To evaluate the effectiveness of image anonymization, we perform linkage attacks on anonymized data and analyze the respective success rates. Furthermore, to quantify the extent of data utility preservation despite the induced image deformations, we compare the performance of a trained thoracic abnormality classification system on anonymized images versus the performance on real data. Throughout our study, we utilize Privacy-Net~\cite{kim2021privacy} and DP-Pix~\cite{fan2018image,fan2019differential} as baseline obfuscation methods.

\begin{figure}[tb]
    \centering
    \includegraphics[width=\textwidth]{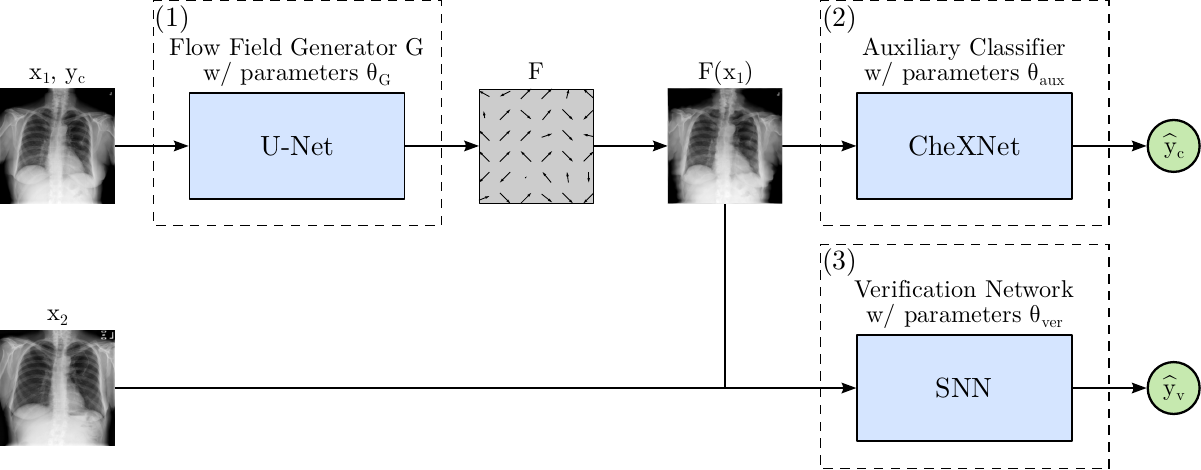}
    \caption{{\bf PriCheXy-Net:} Proposed adversarial image anonymization architecture.}
    \label{fig:network_architecture}
\end{figure}

\section{Methods}
\subsection{Data}
We use X-ray data from the ChestX-ray14 dataset~\cite{wang2017chestx}, a large-scale collection of $112{,}120$ frontal-view chest radiographs from $30{,}805$ unique patients. The 8-bit gray-scale images are provided with a resolution of 1024$\times$1024 pixels. We resize the images to a size of 256$\times$256 pixels for further processing. On average, the dataset contains $\approx$\,3.6 images per patient. The corresponding 14 abnormality labels include Atelectasis, Cardiomegaly, Consolidation, Edema, Effusion, Emphysema, Fibrosis, Hernia, Infiltration, Mass, Nodule, Pleural Thickening, Pneumonia, and Pneumothorax. A patient-wise splitting strategy is applied to roughly divide the data into a training, validation, and test set by ratio 70:10:20, respectively. Based on this split, we utilize available follow-up scans to randomly construct positive image pairs (two unique images from the same patient) and negative image pairs (two unique images from two different patients) -- 10,000 for training, 2,000 for validation, and 5,000 for testing. The resulting subsets are balanced with respect to the number of positive and negative samples.

\subsection{PriCheXy-Net: Adversarial Image Anonymization}
The proposed adversarial image anonymization approach is depicted in Fig.~\ref{fig:network_architecture}. It is composed of three trainable components: (1)~A U-Net generator~$G$ that predicts a flow field~$F$ used to deform the original image~$x_1$ of abnormality class~$y_c$, (2)~an auxiliary classifier that takes the modified image~$F(x_1)$ resulting in corresponding class  predictions~$\hat{y}_c$, and (3)~a siamese neural network~(SNN) that receives the deformed image~$F(x_1)$ as well as another real image~$x_2$ of either the same or a different patient and yields the similarity score~$\hat{y}_v$ for patient verification.
The U-Net serves as an anonymization tool aiming to obfuscate biometric information through targeted image deformations, while the auxiliary classifier and the patient verification model contribute as guidance to optimize the flow field~generator.

\subsubsection{U-Net.}
The U-Net~\cite{ronneberger2015u} architecture is implemented according to Buda~et~al.~\cite{buda2019association}. Its last sigmoid activation function is replaced by a hyperbolic tangent activation function to predict a 2-channel flow field $F$, bounded by~$[-1,1]$. During training, especially in early stages, the raw output of the U-Net may lead to random deformations that destroy the content of the original images, thus revoking the diagnostic utility. To circumvent this issue, the following constraints are imposed for $F$. First, to ensure that the learned flow field~$F$ does not substantially deviate from the identity~$F_{id}$, it is weighted with factor~$\mu$ and subsequently subtracted from the identity according to
\begin{equation}
    F = F_{id} - \mu F\enspace.
    \label{eq:constraints}
\end{equation}
Factor $\mu$ controls the degree of deformation, with larger values allowing for more deformation. Note that the exclusive use of $F_{id}$ would result in the original image, i.\,e., $x = F(x)$, assuming the deformation factor is being set to~$\mu = 0$. The resulting flow field~$F$ is Gaussian filtered (kernel size $9$, $\sigma = 2$) to ensure smooth deformations in the final image. The corresponding parameters were selected manually in preliminary experiments.

\subsubsection{Auxiliary Classifier.}
To ensure the preservation of underlying abnormality patterns and image utility during deformation, PriCheXy-Net integrates an auxiliary classifier using CheXNet~\cite{rajpurkar2017chexnet}, a densely connected convolutional network~(DenseNet)~\cite{huang2017densely} consisting of 121 layers. It outputs a 14-dimensional probability vector~$\hat{y}_c$ indicating the presence or absence of each abnormality appearing in the ChestX-ray14 dataset. Its parameters~$\theta_{aux}$ are initialized using a pre-trained model that achieves a mean AUC of 80.5\,\%.

\subsubsection{Patient Verification Network.}
The incorporated patient verification model is represented by the SNN architecture presented by Packhäuser et al.~\cite{packhauser2022deep}, consisting of two ResNet-50~\cite{he2016deep} branches that are merged using the absolute difference of their resulting 128-dimensional feature vectors. A fully-connected layer with a sigmoid activation function produces the final verification score~$\hat{y}_v$ in the value range of $[0, 1]$, indicating the probability of whether or not the two input images belong to the same patient. For initializing the network parameters~$\theta_{ver}$, we employ a pre-trained network that has been created according to~\cite{packhauser2022deep} yielding an AUC value of $99.4$\,\% for a patient verification task.

\subsection{Objective Functions}
Similar to most adversarial models, our system undergoes training through the use of dual loss functions that guide the model towards opposing directions. To enforce the U-Net not to eliminate important class information while deforming a chest radiograph, we introduce the auxiliary classifier loss $L_{aux}(\theta_G, \theta_{aux})$ realized by the class-wise binary cross entropy~(BCE) loss according to Eq.~\ref{eq:aux_loss}
\begin{equation}
    L_{aux}(\theta_G, \theta_{aux}) = - \sum_{i=1}^{14} [y_{c,i} \log (\hat{y}_{c,i}) + (1-y_{c,i}) \log (1 - \hat{y}_{c,i})]\enspace,
    \label{eq:aux_loss}
\end{equation}
where $i$ represents one out of 14 abnormality classes. Conversely, to guide the U-Net with deceiving the incorporated patient verification model, we utilize its output as an additional verification loss term $L_{ver}(\theta_G, \theta_{ver})$ (see Eq.~\ref{eq:ver_loss}):
\begin{equation}
    L_{ver}(\theta_G, \theta_{ver}) = - \log (1 - \hat{y}_{v})\enspace,\enspace \text{with } \hat{y}_{v} = \text{SNN}\big(F(x_1), x_2\big)\enspace
    \label{eq:ver_loss}
\end{equation}
The total loss to be minimized (see Eq.~\ref{eq:total_loss}) results from the sum of the two partial losses $L_{aux}(\theta_G, \theta_{aux})$ and $L_{ver}(\theta_G, \theta_{ver})$:
\begin{equation}
    \argmin_{\theta_G} L(\theta_G, \theta_{aux}, \theta_{ver}) = L_{aux}(\theta_G, \theta_{aux}) + L_{ver}(\theta_G, \theta_{ver})
    \label{eq:total_loss}
\end{equation}
Lastly, both the auxiliary classifier and the verification model are updated by minimizing the loss terms in Eq.~\ref{eq:aux} and Eq.~\ref{eq:ver}, respectively. Note that the similarity labels for positive and negative pairs are encoded using~$y_v = 1$ and ~$y_v = 0$.
\noindent\begin{minipage}{.35\linewidth}
\begin{equation}
  \argmin_{\theta_{aux}} L(\theta_G, \theta_{aux})
  \label{eq:aux}
\end{equation}
\end{minipage}%
\begin{minipage}{.65\linewidth}
\begin{equation}
  \argmin_{\theta_{ver}} \left[- y_v \log \hat{y}_{v} - (1 - y_v) \log (1 - \hat{y}_{v})\right]
  \label{eq:ver}
\end{equation}
\end{minipage}

\section{Experiments and Results}
\subsection{Experimental Setup}
For all experiments, we used  PyTorch~(1.10.2)~\cite{paszke2019pytorch} and Python~(3.9.5). We followed a multi-part experimental setup consisting of the following steps.

\subsubsection{Pre-training of the flow field generator.}
The incorporated U-Net architecture was pre-trained on an autoencoder-like reconstruction task for 200 epochs using the mean squared error~(MSE) loss,  Adam~\cite{kingma2014adam}, a batch size of 64 and a learning rate of $10^{-4}$ to enable faster convergence. The model that performed best on the validation set was then used for weight initialization in step~2.

\subsubsection{Training of PriCheXy-Net.}
After pre-training, PriCheXy-Net was trained in an end-to-end fashion for 250 epochs using the Adam optimizer~\cite{kingma2014adam}, a batch size of 64 and a learning rate of $10^{-4}$ using the objective functions presented above. To evaluate the effect of the deformation degree $\mu$ on anonymization capability and image utility, we performed multiple training runs with various values~$\mu \in \{0.001, 0.005, 0.01\}$. For each configuration, the U-Net that performed best on the validation set was then used for further evaluations in steps 3 and 4.

\subsubsection{Re-training and evaluation of the verification model.}
To assess the anonymization capability of PriCheXy-Net and to determine if the anonymized images can reliably deceive the verification model, we re-trained the incorporated SNN for each model configuration by using deformed images only. We then simulated multiple linkage attacks by comparing deformed images with real ones. Training was conducted until early stopping~(patience~$p=5$) using the BCE loss, the Adam optimizer~\cite{kingma2014adam}, a batch size of~32 and a learning rate of~$10^{-4}$. For each model configuration, we performed 10 independent training and testing runs. We report the means and standard deviations of the resulting AUC values.

\subsubsection{Evaluation of the classification model on anonymized data.}
To assess the extent to which underlying abnormalities, and thus data utility, were preserved during the anonymization process, each individually trained anonymization network was used to perturb the images of our test set. Then, the pre-trained auxiliary classifier was evaluated using the resulting images. We report the mean of the 14 class-wise AUC values. To quantify the uncertainty, the 95\,\% confidence intervals~(CIs) from $1{,}000$ bootstrap runs were computed.

\subsubsection{Comparison with other obfuscation-based methods.}
To compare our proposed system with other obfuscation-based methods, we additionally analyzed the anonymization capability and utility preservation of Privacy-Net~\cite{kim2021privacy} and DP-Pix~\cite{fan2018image,fan2019differential}. Since Privacy-Net was originally proposed for segmentation tasks, we replaced its segmentation component with the auxiliary classifier. Then, the network was trained and evaluated in the exact same setting as PriCheXy-Net. For DP-Pix, we investigated the effect of different cell sizes~$b \in \{1, 2, 4, 8\}$ at a common privacy budget~$\epsilon = 0.1$. The $m$-neighborhood was set to the smallest possible value~($m=1$) to prevent the added Laplace noise (mean: 0; scale: $\frac{255m}{b^2\epsilon}$) from destroying the complete content of the images.
\begin{table}[H]
    \centering
    \caption{Quantitative comparison of all examined methods. The baseline performance results from leveraging non-anonymized real data. Verification: AUC (mean $\pm$ std) over 10 independent training and testing runs; Classification: mean AUC + 95\,\% CIs. Performance scores of 50\,\% indicate random decisions.}
    \input{paper_table1}
    \label{tab:results_combined}
\end{table}

\subsection{Results}
\subsubsection{Baseline and comparison methods.}
The results of all conducted experiments are shown in Table~\ref{tab:results_combined}. Compared to real (non-anonymized) data, which enables a successful re-identification with an AUC of 81.8\,\%, the patient verification performance desirably decreases after applying DP-Pix~(50.0\,\% - 52.5\,\%) and Privacy-Net~(49.8\,\%). However, while the classification performance on real data indicates a high data utility with a mean AUC of 80.5\,\%, we observe a sharp drop for images that have been modified with DP-Pix~(50.0\,\% - 52.9\,\%) and Privacy-Net~(57.5\,\%). This suggests that relevant class information and specific abnormality patterns are destroyed during the obfuscation process. Resulting example images for both comparison methods are provided in~Suppl.~Fig.~\ref{fig:suppl1}.
\begin{figure}[tb]
    \centering
    \includegraphics[width=0.44\textwidth]{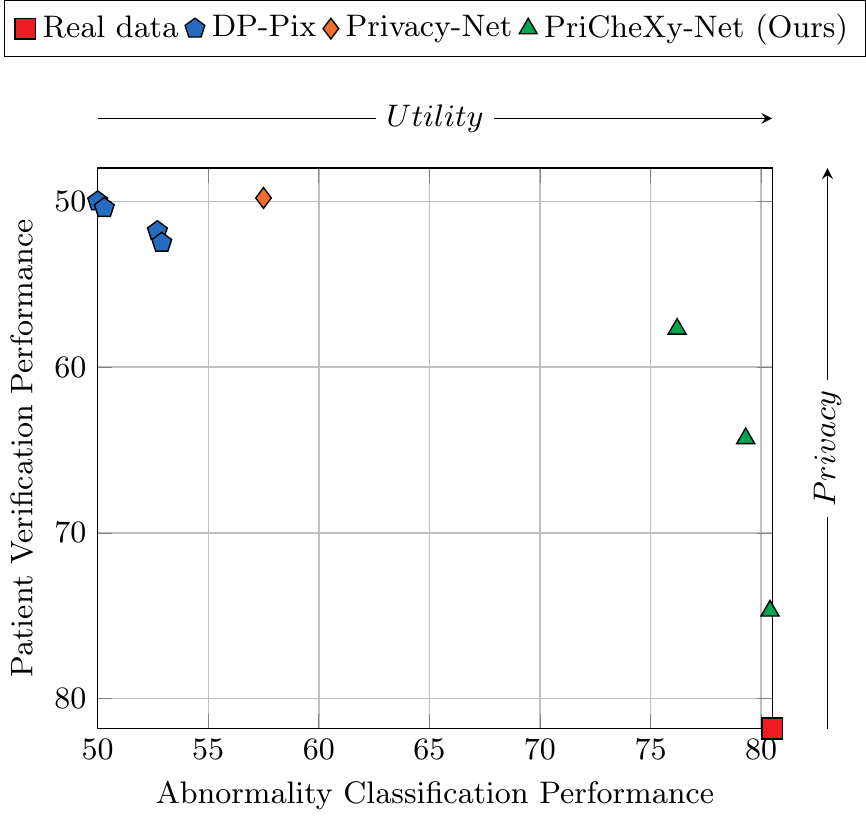}
    \caption{Visual illustration of the results and the associated privacy-utility trade-off. The patient verification performance (y-axis) measures the amount of privacy, whereas the abnormality classification performance (x-axis) represents the level of data utility.}
    \label{fig:tradeoff_results_train_loss_networks}
\end{figure}
\begin{figure}[tb]
  \centering
  \subfigure[real\label{fig:1}]{\includegraphics[width=.22\textwidth]{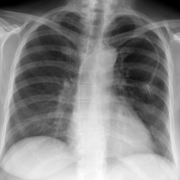}} \quad
  \subfigure[$\mu=0.001$\label{fig:2}]{\includegraphics[width=.22\textwidth]{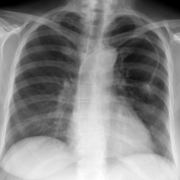}} \quad
  \subfigure[$\mu=0.005$\label{fig:3}]{\includegraphics[width=.22\textwidth]{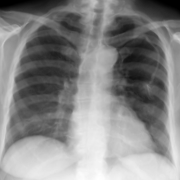}} \quad
  \subfigure[$\mu=0.01$\label{fig:4}]{\includegraphics[width=.22\textwidth]{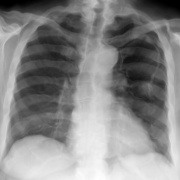}}
  \caption{Chest radiographs resulting from PriCheXy-Net when using different deformation degrees~$\mu$. Images were cropped to better highlight the diagnostically relevant~area.}\label{fig:images}
\end{figure}

\subsubsection{PriCheXy-Net.}
The results of our proposed PriCheXy-Net~(see Table~\ref{tab:results_combined}) show more promising behavior. As can be seen, increasing deformation degrees~$\mu$ lead to a successive decline in patient verification performance. Compared to the baseline, the AUC decreases to 74.7\,\%~($\mu=0.001$), to 64.3\,\%~($\mu=0.005$), and to 57.7\,\%~($\mu=0.01$), indicating a positive effect on patient privacy and the obfuscation of biometric information. In addition, PriCheXy-Net hardly results in a loss of data utility, as characterized by the constantly high classification performance with a mean AUC of 80.4\,\%~($\mu=0.001$), 79.3\,\%~($\mu=0.005$), and 76.2\,\%~($\mu=0.01$). These findings are further visualized in the privacy-utility trade-off plot in Fig.~\ref{fig:tradeoff_results_train_loss_networks}. In contrast to the examined comparison methods, the data point corresponding to our best experiment with PriCheXy-Net~(green) lies near the top right corner, highlighting the capability to closely satisfy both objectives in the privacy-utility trade-off. Examples of deformed chest radiographs resulting from a trained model of PriCheXy-Net are shown in Fig.~\ref{fig:images}. More examples are given in Suppl.~Fig.~\ref{fig:suppl1}. Difference maps are provided in Suppl.~Fig.~\ref{fig:suppl2}.

\section{Discussion and Conclusion}
To the best of our knowledge, we presented the first adversarial approach to anonymize thoracic images while preserving data utility for diagnostic purposes. Our proposed anonymization approach -- PriCheXy-Net -- is a composition of three independent neural networks consisting of (1)~a flow field generator, (2)~an auxiliary classifier, and (3)~a patient verification network. In this work, we were able to show that collective utilization of these three components enables learning of a flow field that targetedly deforms chest radiographs and thus reliably deceives a patient verification model, even after re-training was performed. For the best hyper-parameter configuration of PriCheXy-Net, the re-identification performance drops from~$81.8$\,\% to~$57.7$\,\% in AUC for a simulated linkage attack, whereas the abnormality classification performance only decreases from $80.5$\,\% to~$76.2$\,\%, which indicates the effectiveness of the proposed approach. We strongly hypothesize that the promising performance of PriCheXy-Net can be largely attributed to the constraints imposed on the learned flow field~$F$. The limited deviation of the flow field from the identity~(cf.~Eq.~\ref{eq:constraints}) ensures a realistic appearance of the resulting deformed image to a considerable extent, thereby avoiding its content from being completely destroyed. This idea has a positive impact on preserving relevant abnormality patterns in chest radiographs, while allowing adequate scope to obfuscate biometric information. Such domain-specific constraints are not integrated in examined comparison methods such as Privacy-Net (which directly predicts an anonymized image without ensuring realism) and DP-Pix (which does not contain any mechanism to maintain data utility). This is, as we hypothesize, the primary reason for their limited ability to preserve data utility and the overall superiority of our proposed system. Interestingly, PriCheXy-Net's deformation fields primarily focus on anatomical structures, including lungs and ribs, as demonstrated in Fig.~\ref{fig:images}, Suppl.~Fig.~\ref{fig:suppl1}, and Suppl.~Fig.~\ref{fig:suppl2}. This observation aligns with Packhäuser et al.'s previous findings~\cite{packhauser2022deep}, which revealed that these structures contain the principal biometric information in chest radiographs.

In future work, we aim to further improve the performance of PriCheXy-Net by incorporating additional components into its current architecture. For instance, we plan to integrate a discriminator loss into the model, which may positively contribute to achieving perceptual realism. Furthermore, we also consider implementing a region of interest segmentation step into the pipeline to ensure not to perturb diagnostically relevant image areas. Lastly, we hypothesize that our method is robust to variations in image size or compression rate, and posit its applicability beyond chest X-rays to other imaging modalities as well. However, confirmation of these hypotheses requires further exploration to be conducted in forthcoming studies.

\subsubsection{Data use declaration.}
This research study was conducted retrospectively using human subject data made available in open access by the National Institutes of Health (NIH) Clinical Center~\cite{wang2017chestx}. Ethical approval was not required as confirmed by the license attached with the open-access data.

\subsubsection{Code availability.}
The source code of this study has been made available at \url{https://github.com/kaipackhaeuser/PriCheXy-Net}.

\subsubsection{Acknowledgments.}
The research leading to these results has received funding from the European Research Council (ERC) under the European Union’s Horizon 2020 research and innovation program (ERC Grant no. 810316). 
The authors gratefully acknowledge the scientific support and HPC resources provided by the Erlangen National High Performance Computing Center (NHR@FAU) of the Friedrich-Alexander-Universität Erlangen-Nürnberg (FAU). The hardware is funded by the German Research Foundation (DFG). The authors declare that they have no conflicts of interest.

%
%
%
\bibliographystyle{splncs04}
\bibliography{paper}

\renewcommand{\figurename}{Suppl.~Fig.}
\setcounter{figure}{0}
\renewcommand\thefigure{\arabic{figure}}
\renewcommand{\theHfigure}{\thefigure}

\begin{figure}
    \centering
    \includegraphics[width=.91\textwidth]{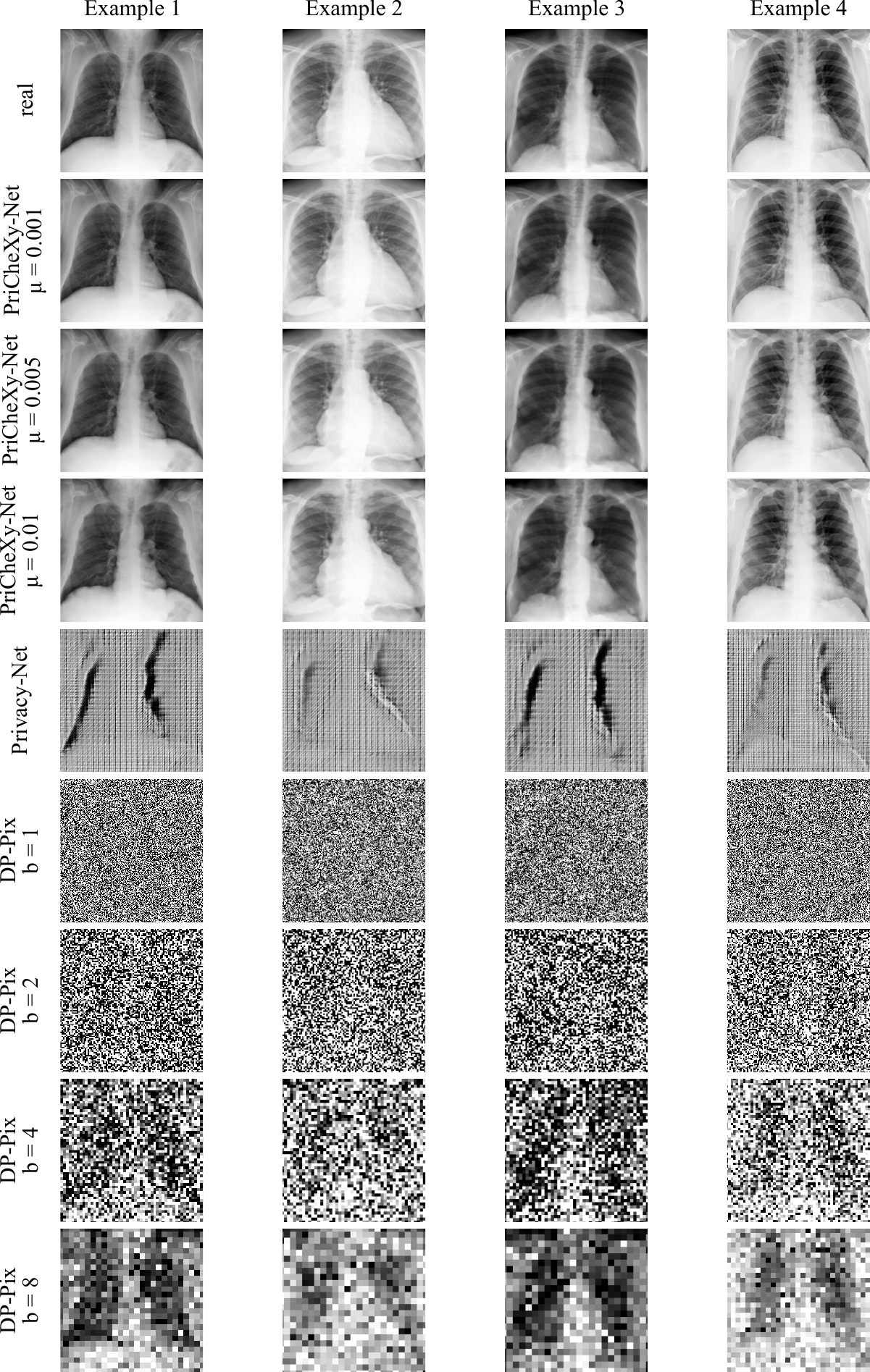}
    \caption{Example images resulting from different obfuscation-based anonymization techniques at various hyper-parameter configurations. Row 1: Real chest X-rays. Rows 2-4: PriCheXy-Net (ours). Row 5: Privacy-Net. Rows 6-9: DP-Pix.}
    \label{fig:suppl1}
\end{figure}

\begin{figure}
    \centering
    \includegraphics[width=.91\textwidth]{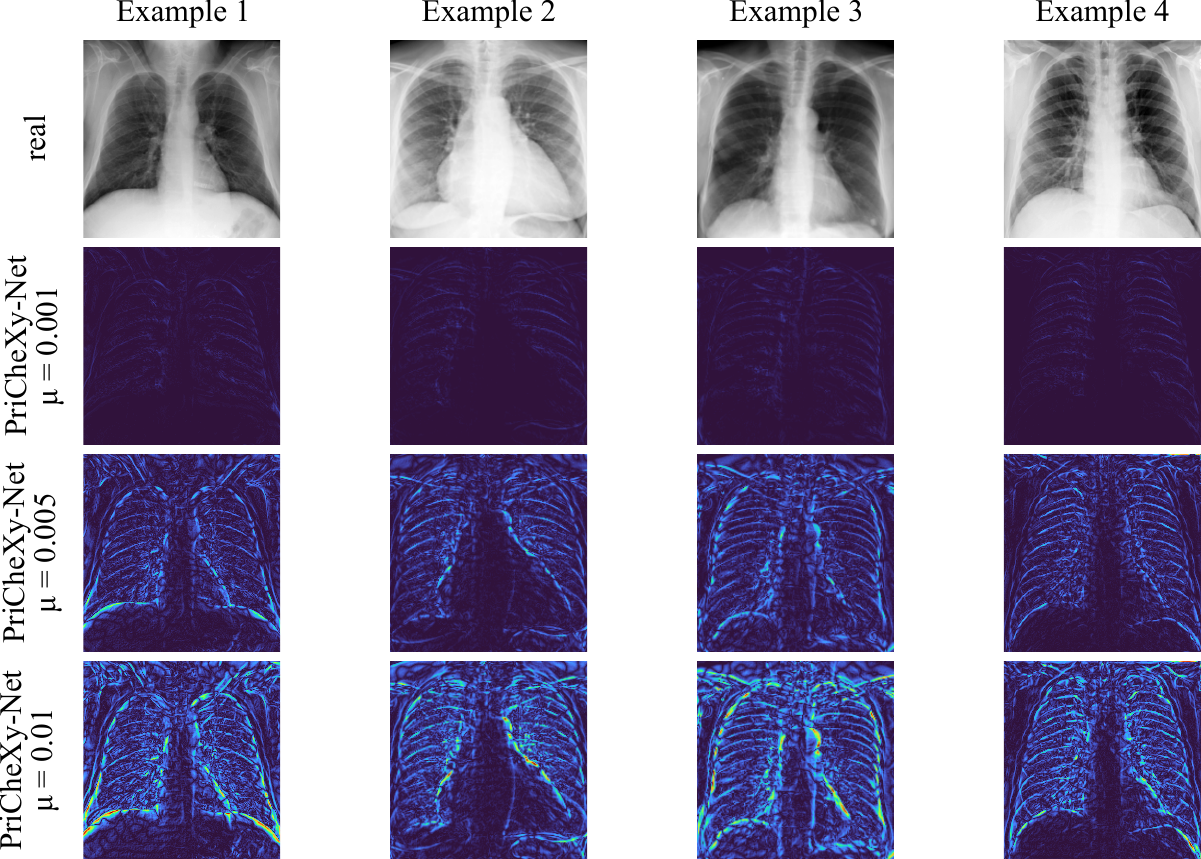}
    \caption{Visualizations of difference maps computed according to $|x - F(x)|$\,, with $F$ being the learned deformation field that is applied to an image $x$. As shown, PriCheXy-Net targets anatomical structures, such as lungs and ribs, which contain the principal biometric information in chest radiographs.}
    \label{fig:suppl2}
\end{figure}

\begin{figure}
\centering

\begin{minipage}[t]{0.49\textwidth}
\centering
    \begin{algorithm}[H]
    \SetKwComment{Comment}{/* }{ */}
    \SetKwInOut{Input}{Input}\SetKwInOut{Output}{Output}
    \caption{Training procedure.}\label{alg:training}
    \Input{Images $\mathcal{X}_1$, \\Images $\mathcal{X}_2$, \\Class labels $\mathcal{Y}_c$, \\Similarity labels $\mathcal{Y}_v$}
    \BlankLine
    \Output{Network parameters $\theta_G$}
    \BlankLine
    \Comment{Network initialization}
    Initialize network parameters $\theta_G$, $\theta_{aux}$, $\theta_{ver}$\;
    \BlankLine
    \Comment{Main loop}
    \For{$\text{epoch}=1, ..., E_{\text{max}}$}{
      \For{$\text{iter}=1, ..., I_{\text{max}}$}{
        \Comment{Random batch selection}
        Randomly select batch $\mathcal{B}$\;
        \BlankLine
        \Comment{Update flow field generator}
        Update $\theta_G$ according to Eq.~4\;
        \BlankLine
        \Comment{Update auxiliary classifier}
        Update $\theta_{aux}$ according to Eq.~5\;
        \BlankLine
        \Comment{Update verification network}
        Update $\theta_{ver}$ according to Eq.~6\;
      }
    }
    \BlankLine
    \KwRet{$\theta_G$\;}
    \end{algorithm}
\end{minipage}
\hfill
\begin{minipage}[t]{0.49\textwidth}
\centering
    \begin{algorithm}[H]
    \SetKwComment{Comment}{/* }{ */}
    \SetKwInOut{Input}{Input}\SetKwInOut{Output}{Output}
    \caption{Inference procedure.}\label{alg:inference}
    \Input{Image $x$, \\Network parameters $\theta_G$}
    \BlankLine
    \Output{Anonymized image $F(x)$}
    \BlankLine
    \Comment{Network initialization}
    Initialize flow field generator $G$ with learned network parameters $\theta_G$\;
    \BlankLine
    \Comment{Flow field prediction}
    Predict the flow field $F$ for image $x$ by applying the flow field generator $G$\;
    \BlankLine
    \Comment{Image anonymization}
    Generate anonymized image $F(x)$ by applying the predicted flow field $F$ to image $x$\;
    \BlankLine
    \KwRet{$F(x)$\;}
    \end{algorithm}
\end{minipage}
\caption{Pseudocode that outlines the training procedure (Algorithm 1) and the inference procedure (Algorithm 2), respectively.}
\end{figure}

\end{document}

%% file: paper_table1.tex
\newcolumntype{Y}{>{\centering\arraybackslash}X}
\begin{tabularx}{\textwidth}{lYYYYY}
\toprule
    &   \textbf{Baseline} & \textbf{Privacy-Net} & \multicolumn{3}{c}{\textbf{PriCheXy-Net (Ours)}} \\ \cmidrule{4-6}
    \textbf{Task} & (real data) & \cite{kim2021privacy} & $\mu=0.001$ & $\mu=0.005$ & $\mu=0.01$ \\ \midrule
    Ver. $\downarrow$    & $81.8 \pm 0.6$ & $49.8 \pm 2.2$ & $74.7 \pm 2.6$     & $64.3 \pm 5.8$     & $57.7 \pm 4.0$    \\
    Class. $\uparrow$ & $80.5\enspace_{80.1}^{80.9}$ & $57.5\enspace_{56.9}^{58.1}$ & $80.4\enspace_{79.9}^{80.8}$     & $79.3\enspace_{78.9}^{79.7}$     & $76.2\enspace_{75.8}^{76.6}$    \\\midrule \midrule

    &   &   \multicolumn{4}{c}{\textbf{DP-Pix}~\cite{fan2018image,fan2019differential}}                 \\ \cmidrule{3-6}
    &  & $b=1$ & $b=2$ & $b=4$ & $b=8$ \\ \midrule
    Ver. $\downarrow$    & $81.8 \pm 0.6$ & $50.0 \pm 0.7$ & $50.4 \pm 0.6$     & $51.8 \pm 1.3$     & $52.5 \pm 3.2$    \\
    Class. $\uparrow$ & $80.5\enspace_{80.1}^{80.9}$ & $50.0\enspace_{49.4}^{50.6}$ & $50.3\enspace_{49.7}^{51.0}$     & $52.7\enspace_{52.0}^{53.3}$     & $52.9\enspace_{52.2}^{53.5}$    \\\bottomrule
\end{tabularx}